\documentclass[manuscript]{aastex}

\usepackage{lineno}

\usepackage{esvect}

\usepackage{gensymb}

\usepackage{multirow}

\usepackage{amsmath}

\begin{document}


\title{Search for TeV Gamma-Ray Emission from Point-like Sources in the Inner Galactic Plane with a Partial Configuration of the HAWC Observatory}

\renewcommand*{\thefootnote}{\alph{footnote}}
\author{A.~U.~Abeysekara\altaffilmark{1,2},
R.~Alfaro\altaffilmark{3,4},
C.~Alvarez\altaffilmark{5},
J.~D.~{\'A}lvarez\altaffilmark{6},
R.~Arceo\altaffilmark{5},
J.~C.~Arteaga-Vel{\'a}zquez\altaffilmark{6},
H.~A.~Ayala~Solares\altaffilmark{7},
A.~S.~Barber\altaffilmark{2},
B.~M.~Baughman\altaffilmark{4},
N.~Bautista-Elivar\altaffilmark{8},
A.~D.~Becerril~Reyes\altaffilmark{3},
E.~Belmont\altaffilmark{3},
S.~Y.~BenZvi\altaffilmark{9,10},
A.~Bernal\altaffilmark{11},
J.~Braun\altaffilmark{4,10},
K.~S.~Caballero-Mora\altaffilmark{12},
T.~Capistr{\'a}n\altaffilmark{13},
A.~Carrami{\~n}ana\altaffilmark{13},
S.~Casanova\altaffilmark{16},
M.~Castillo\altaffilmark{6},
U.~Cotti\altaffilmark{6},
J.~Cotzomi\altaffilmark{14},
S.~Couti{\~n}o~de~Le{\'o}n\altaffilmark{13},
E.~de~la~Fuente\altaffilmark{15},
C.~De~Le{\'o}n\altaffilmark{6},
T.~DeYoung\altaffilmark{1},
B.~L.~Dingus\altaffilmark{17},
M.~A.~DuVernois\altaffilmark{10},
R.~W.~Ellsworth\altaffilmark{4,18},
O.~Enriquez-Rivera\altaffilmark{21},
D.~W.~Fiorino\altaffilmark{10},
N.~Fraija\altaffilmark{11},
F.~Garfias\altaffilmark{11},
M.~M.~Gonz{\'a}lez\altaffilmark{4,11},
J.~A.~Goodman\altaffilmark{4},
M.~Gussert\altaffilmark{19},
Z.~Hampel-Arias\altaffilmark{10},
J.~P.~Harding\altaffilmark{17},
S.~Hernandez\altaffilmark{3},
P.~H{\"u}ntemeyer\altaffilmark{7},
C.~M.~Hui\altaffilmark{7},
A.~Imran\altaffilmark{10,17,}\footnotemark[1] \footnotetext[1]{current address: AdRoll Inc., San Francisco, CA, USA},
A.~Iriarte\altaffilmark{11},
P.~Karn\altaffilmark{10,20},
D.~Kieda\altaffilmark{2},
A.~Lara\altaffilmark{21},
R.~J.~Lauer\altaffilmark{22},
W.~H.~Lee\altaffilmark{11},
D.~Lennarz\altaffilmark{23},
H.~Le{\'o}n~Vargas\altaffilmark{3},
J.~T.~Linnemann\altaffilmark{1},
M.~Longo\altaffilmark{19},
G.~Luis~Raya\altaffilmark{8},
K.~Malone\altaffilmark{25},
A.~Marinelli\altaffilmark{3},
S.~S.~Marinelli\altaffilmark{1},
H.~Martinez\altaffilmark{12},
O.~Martinez\altaffilmark{14},
J.~Mart{\'\i}nez-Castro\altaffilmark{24},
J.~A.~Matthews\altaffilmark{22},
P.~Miranda-Romagnoli\altaffilmark{26},
E.~Moreno\altaffilmark{14},
M.~Mostaf{\'a}\altaffilmark{25},
L.~Nellen\altaffilmark{27},
M.~Newbold\altaffilmark{2},
R.~Noriega-Papaqui\altaffilmark{26},
B.~Patricelli\altaffilmark{11},
R.~Pelayo\altaffilmark{24},
E.~G.~P{\'e}rez-P{\'e}rez\altaffilmark{8},
J.~Pretz\altaffilmark{25},
Z.~Ren\altaffilmark{22},
C.~Rivi\`ere\altaffilmark{4,11},
D.~Rosa-Gonz{\'a}lez\altaffilmark{13},
H.~Salazar\altaffilmark{14},
F.~Salesa~Greus\altaffilmark{25},
A.~Sandoval\altaffilmark{3},
M.~Schneider\altaffilmark{28},
G.~Sinnis\altaffilmark{17},
A.~J.~Smith\altaffilmark{4},
K.~Sparks~Woodle\altaffilmark{25},
R.~W.~Springer\altaffilmark{2},
I.~Taboada\altaffilmark{23},
O.~Tibolla\altaffilmark{5},
K.~Tollefson\altaffilmark{1},
I.~Torres\altaffilmark{13},
T.~N.~Ukwatta\altaffilmark{1,17},
L.~Villase{\~n}or\altaffilmark{6},
K.~Vrabel\altaffilmark{25},
T.~Weisgarber\altaffilmark{10},
S.~Westerhoff\altaffilmark{10},
I.~G.~Wisher\altaffilmark{10},
J.~Wood\altaffilmark{4},
T.~Yapici\altaffilmark{1},
G.~B.~Yodh\altaffilmark{20},
P.~W.~Younk\altaffilmark{17},
D.~Zaborov\altaffilmark{25,}\footnotemark[2] \footnotetext[2]{current address: LLR - Ecole polytechnique, CNRS/IN2P3, Palaiseau, France},
A.~Zepeda\altaffilmark{12},
and H.~Zhou\altaffilmark{7}
\newline {(The HAWC collaboration)}}

\altaffiltext{1}{Department of Physics \& Astronomy, Michigan State University, East Lansing, MI, USA}
\altaffiltext{2}{Department of Physics \& Astronomy, University of Utah, Salt Lake City, UT, USA}
\altaffiltext{3}{Instituto de F{\'\i}sica, Universidad Nacional Aut{\'o}noma de M{\'e}xico, M{\'e}xico D.F., Mexico}
\altaffiltext{4}{Department of Physics, University of Maryland, College Park, MD, USA}
\altaffiltext{5}{Universidad Aut{\'o}noma de Chiapas, Tuxtla Guti{\'e}rrez, Chiapas, Mexico}
\altaffiltext{6}{Universidad Michoacana de San Nicol{\'a}s de Hidalgo, Morelia, Michoac{\'a}n, Mexico}
\altaffiltext{7}{Department of Physics, Michigan Technological University, Houghton, MI, USA}
\altaffiltext{8}{Universidad Polit{\'e}cnica de Pachuca, Municipio de Zempoala, Hidalgo, Mexico}
\altaffiltext{9}{Department of Physics \& Astronomy, University of Rochester, Rochester, NY, USA}
\altaffiltext{10}{Wisconsin IceCube Particle Astrophysics Center (WIPAC) and Department of Physics, University of Wisconsin-Madison, Madison, WI, USA}
\altaffiltext{11}{Instituto de Astronom{\'\i}a, Universidad Nacional Aut{\'o}noma de M{\'e}xico, M{\'e}xico D.F., Mexico}
\altaffiltext{12}{Centro de Investigaci{\'o}n y de Estudios Avanzados del Instituto Polit{\'e}cnico Nacional, M{\'e}xico D.F., Mexico}
\altaffiltext{13}{Instituto Nacional de Astrof{\'\i}sica, {\'O}ptica y Electr{\'o}nica, Tonantzintla, Puebla, Mexico}
\altaffiltext{14}{Facultad de Ciencias F{\'\i}sico Matem{\'a}ticas, Benem{\'e}rita Universidad Aut{\'o}noma de Puebla, Ciudad Universitaria, Puebla, Mexico}
\altaffiltext{15}{Departamento de F{\'\i}sica, Centro Universitario de Ciencias Exactas e Ingenier{\'\i}as, Universidad de Guadalajara, Guadalajara, Mexico}
\altaffiltext{16}{Instytut Fizyki Jadrowej im Henryka Niewodniczanskiego Polskiej Akademii Nauk, IFJ-PAN, Krakow, Poland }
\altaffiltext{17}{Physics Division, Los Alamos National Laboratory, Los Alamos, NM, USA}
\altaffiltext{18}{School of Physics, Astronomy \& Computational Sciences, George Mason University, Fairfax, VA, USA}
\altaffiltext{19}{Physics Department, Colorado State University, Fort Collins, CO, USA}
\altaffiltext{20}{Department of Physics \& Astronomy, University of California, Irvine, Irvine, CA, USA}
\altaffiltext{21}{Instituto de Geof{\'\i}sica, Universidad Nacional Aut{\'o}noma de M{\'e}xico, M{\'e}xico D.F., Mexico}
\altaffiltext{22}{Department of Physics \& Astronomy, University of New Mexico, Albuquerque, NM, USA}
\altaffiltext{23}{School of Physics and Center for Relativistic Astrophysics, Georgia Institute of Technology, Atlanta, GA, USA}
\altaffiltext{24}{Centro de Investigaci{\'o}n en Computaci{\'o}n, Instituto Polit{\'e}cnico Nacional, M{\'e}xico D.F., Mexico}
\altaffiltext{25}{Department of Physics, Pennsylvania State University, University Park, PA, USA}
\altaffiltext{26}{Universidad Aut{\'o}noma del Estado de Hidalgo, Pachuca, Hidalgo, Mexico}
\altaffiltext{27}{Instituto de Ciencias Nucleares, Universidad Nacional Aut{\'o}noma de M{\'e}xico, M{\'e}xico D.F., Mexico}
\altaffiltext{28}{Santa Cruz Institute for Particle Physics, University of California, Santa Cruz, Santa Cruz, CA, USA}
\authoraddr{H.~Zhou, \email{hzhou1@mtu.edu}}
\authoraddr{C.~M.~Hui, \email{cmhui@mtu.edu}}



\begin{abstract}
A survey of the inner Galaxy region of Galactic longitude $l\in [+15\degree,\,+50\degree]$ and latitude $b\in[-4\degree,\,+4\degree]$ is performed using one-third of the High Altitude Water Cherenkov (HAWC) Observatory operated during its construction phase.  To address the ambiguities arising from unresolved sources in the data, we use a maximum likelihood technique to identify point source candidates.  Ten sources and candidate sources are identified in this analysis.  Eight of these are associated with known TeV sources but not all have differential fluxes compatible with previous measurements.  Three sources are detected with significances $>5\,\sigma$ after accounting for statistical trials, and are associated with known TeV sources.
\end{abstract}





\section{Introduction}
We present a TeV gamma-ray survey of the inner Galaxy region between +15$\degree$ and +50$\degree$ in Galactic longitude ($l$) and $\pm$4$\degree$ in Galactic latitude ($b$) using data collected with a partially completed configuration of the High Altitude Water Cherenkov (HAWC) Observatory. This region of the sky contains the strongest Galactic detections in the field of view of HAWC other than the Crab Nebula, and is known to have a large number of gamma-ray sources such as pulsars, pulsar wind nebulae (PWNe), supernova remnants (SNRs), compact object binaries, diffuse emission from the Galactic plane, and sources without known astrophysical associations. The most commonly identified TeV Galactic sources are SNRs and PWNe \citep{tevcat}. SNRs are thought to be acceleration sites of Galactic cosmic rays, since they can accelerate particles via diffusive shock acceleration and provide sufficient power to explain the cosmic-ray energy losses from the Galaxy (e.g. \cite{Drury2001}).  In PWNe, electrons effectively gain energy at the termination shock where the pulsar wind is terminated by the surrounding gas, emitting TeV gamma rays via inverse Compton scattering \citep{gammaRay}. Many of the gamma-ray sources in the Galactic plane are unidentified (UID) due to the fact that the measurements can envelope multiple sources identified at other wavelengths.  Morphological and spectral studies are crucial for making associations with observations at other wavelengths and for distinguishing leptonic and hadronic gamma-ray production processes which will aid in source identification.

The HAWC survey presented here overlaps in both energy range and sky coverage with the survey and observations performed by the H.E.S.S. imaging atmospheric Cherenkov telescopes (IACTs) (for example, \cite{HESSsurvey,hessUID,hess1843,hess1846,hess1848}). Over 15 sources within this region were discovered by H.E.S.S.  Similar surveys within this region have been performed by Milagro \citep{milagroSurvey} and ARGO \citep{argoSurvey}, along with targeted observations by the IACTs VERITAS and MAGIC (for example, \citet{magic1834,verj0632,magicW51,veri1908,magic1857}). At lower energies, the \textit{Fermi} Large Area Telescope (\textit{Fermi}-LAT) has published its third source catalog in the 100\,MeV-300\,GeV energy range (3FGL) based on the first four years of science operation \citep{3fgl}. The region surveyed with the partial HAWC array includes the locations of 73 sources from this catalog, 47 of which are without known astronomical associations. The first catalog with sources $>10$\,GeV (1FHL) has been published based on the first three years of \textit{Fermi}-LAT data \citep{1fhl}. The region surveyed in this publication contains twelve sources from this catalog, four of which are not in the 3FGL catalog.

Source identification is a major challenge when analyzing the emission from this region, with point-like and extended emissions overlapping each other. For this reason, a method is applied to the data to simultaneously fit the positions and differential flux normalizations of multiple sources assuming a simple power law with a spectral index of 2.3. The maximum likelihood method used here, similar to the one used by the \textit{Fermi}-LAT \citep{2fgl}, is described in more detail in Section 3, following a description of the HAWC detector in Section 2.  In Section 4, we present a list of 10 sources and candidate sources with locations and differential flux normalizations found with the likelihood method, followed by a discussion of their possible associations with previously reported objects.

\section{The HAWC Gamma-Ray Observatory}
\begin{figure}
\centering
\epsscale{.80}
\plotone{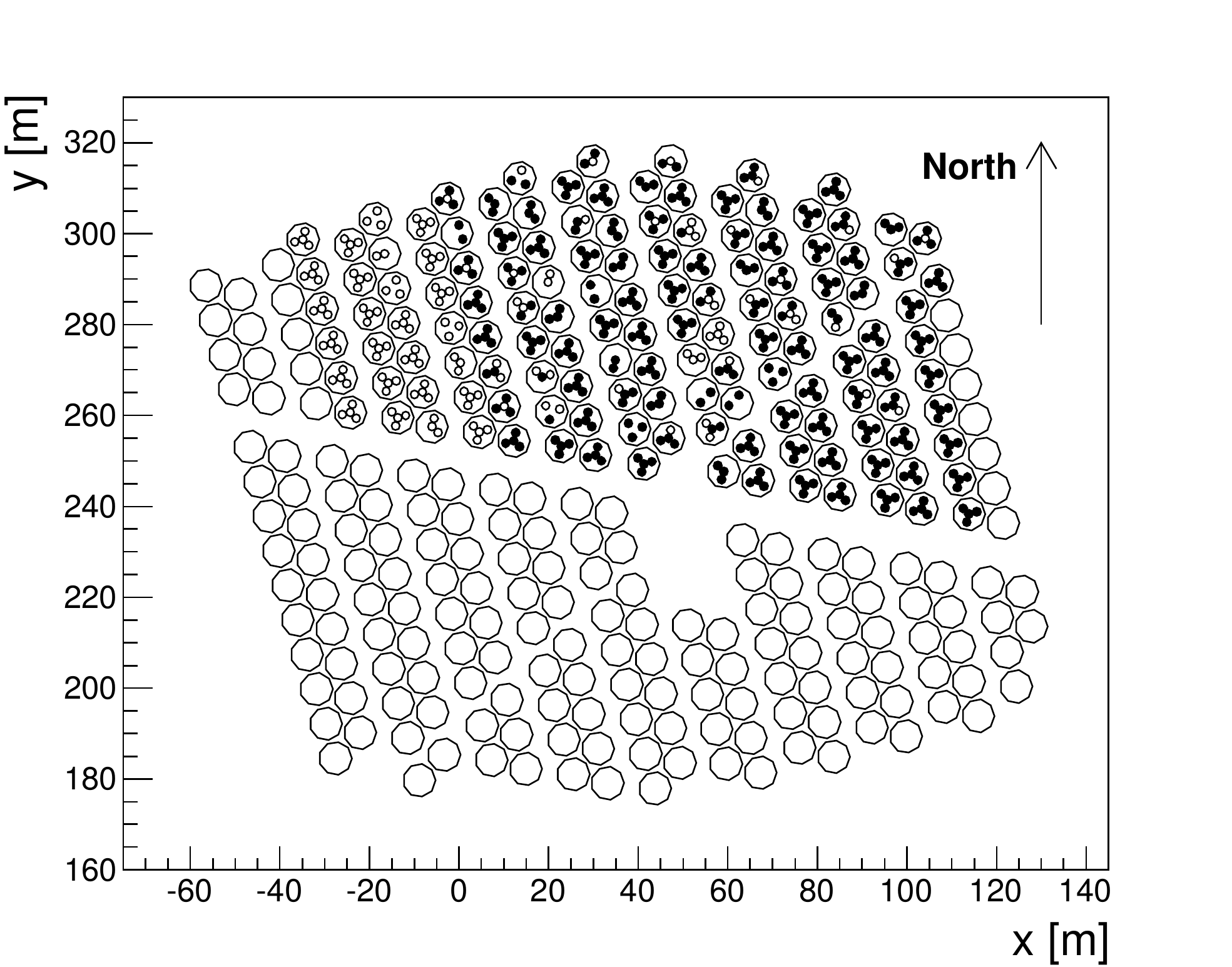}
\caption{The layout of the full HAWC array, with filled circles marking the PMTs operational at the beginning of the dataset under investigation and small open circles indicating the PMTs that were eventually integrated into the array during this period.  WCDs are shown as large circles.  WCDs not commissioned during the period covered by this dataset are drawn without PMTs.}
\label{fig:hawc}
\end{figure}

The HAWC Gamma-Ray Observatory is located at Sierra Negra, Mexico (18$\degree$59'41" N 97$\degree$18'30.6"W) at 4,100\,m a.s.l., and is sensitive to gamma rays and cosmic rays in the energy range from 100\,GeV to 100\,TeV \citep{abeysekara2013a}.  The observatory consists of an array of 300 water Cherenkov detectors (WCDs) as shown in Fig.\,\ref{fig:hawc}, covering an area of 22,000\,$\text{m}^2$.  Each WCD consists of a tank 7.3\,m in diameter, 4.5\,m in depth, and filled with 230,000\,L of purified water.  Four upward-facing photomultiplier tubes (PMTs) are attached to the bottom of each WCD: one high-quantum efficiency 10-inch Hamamatsu R7081-MOD PMT at the center and three 8-inch Hamamatsu R5912 PMTs spaced $120^\circ$ apart 1.8\,m from the center \citep{abeysekara2013b}.  The array has an instantaneous field of view of 2\,sr and a duty cycle $>95$\% is expected once construction is completed.

The PMTs detect Cherenkov light inside the WCDs produced by relativistic secondary particles from an extensive air shower. The PMT signals propagate through 600 feet of RG59 cable to the data acquisition system in a central building, where they are processed by custom-made front end boards. The PMT pulses are shaped and discriminated using a low and a high voltage threshold. The time stamp of a pulse that crosses at least the lower of the two thresholds is recorded by CAEN VX1190A time-to-digital-converter modules with a precision of 0.1\,ns.  The pulse size and ultimately the charge can be inferred from the time-over-threshold (ToT).

A laser calibration system measures the ToT-charge conversion and the response time of each PMT to different pulse sizes \citep{abeysekara2013b}.  A simple multiplicity trigger requiring at least 16 to 21 PMTs above threshold within a 150\,ns time window is used to identify air shower events.  Once an air shower triggers the detector, the charge and timing calibration are applied to each PMT signal.  The Nishimura-Kamata-Greisen (NKG) function \citep{greisen1956,kamata1958} describes the lateral distribution of charged particles produced by an electromagnetic air shower.  The distribution of signal amplitude in the detector is fitted with the NKG function in order to find the position of the shower core, where the extrapolated trajectory of the primary particle intersects the altitude of the detector.  The direction of the primary particle is subsequently reconstructed by taking the estimated position of the shower core and fitting the relative times of the PMT signals produced when the air shower crosses the array.

\section{Data and Analysis}
The HAWC array was completed in March 2015, but science operations began in August 2013.  The analysis described in this paper is performed on data taken with the partially constructed HAWC array (called Pass 1 data/configuration hereafter) between August 2, 2013 and July 9, 2014.  During this period, data taking was occasionally interrupted for detector construction work and maintenance, and the duty cycle is $\sim84\%$.
The dataset contains $275\pm1$ source transits of the inner Galaxy region, depending on the right ascension.  The detector grew 
from 362 PMTs in 108 WCDs to 491 PMTs in 134 WCDs during this time as shown in Fig.\,\ref{fig:hawc}.  The data analyzed are divided into three epochs, distinguished by the number of active PMTs.  

The sensitivity of HAWC is a function of source declination.  The best sensitivity is achieved for sources that transit through the detector zenith.  For the study presented here, a zenith angle cut of $45\degree$ is applied, corresponding to declinations between $-26\degree$ and $+64\degree$.

The energy, angular resolution, and background rejection of the observed gamma rays are correlated with the shower size measured in the array. Therefore, the data are divided into 10 bins according to the fraction \textit{f} of PMTs triggered by an air shower event passing standard selection cuts (e.g. $>1$ PE within 450\,ns time window) out of the total number of active PMTs in the array. Cuts are applied to three parameters in each \textit{f} bin to separate the cosmic-ray background from the gamma-ray signal:
\begin{enumerate}
\item The ratio of the chi-square of the fit to the core location and the number of PMTs in the event; 
\item A topological cut based on the ``compactness'' of the charge distribution in the array;
\item The reduced chi-square of the fit to the shower direction.
\end{enumerate}
The cuts are explained in detail in \citet{Paco2015}.

\begin{figure}
\includegraphics[width=0.5\textwidth]{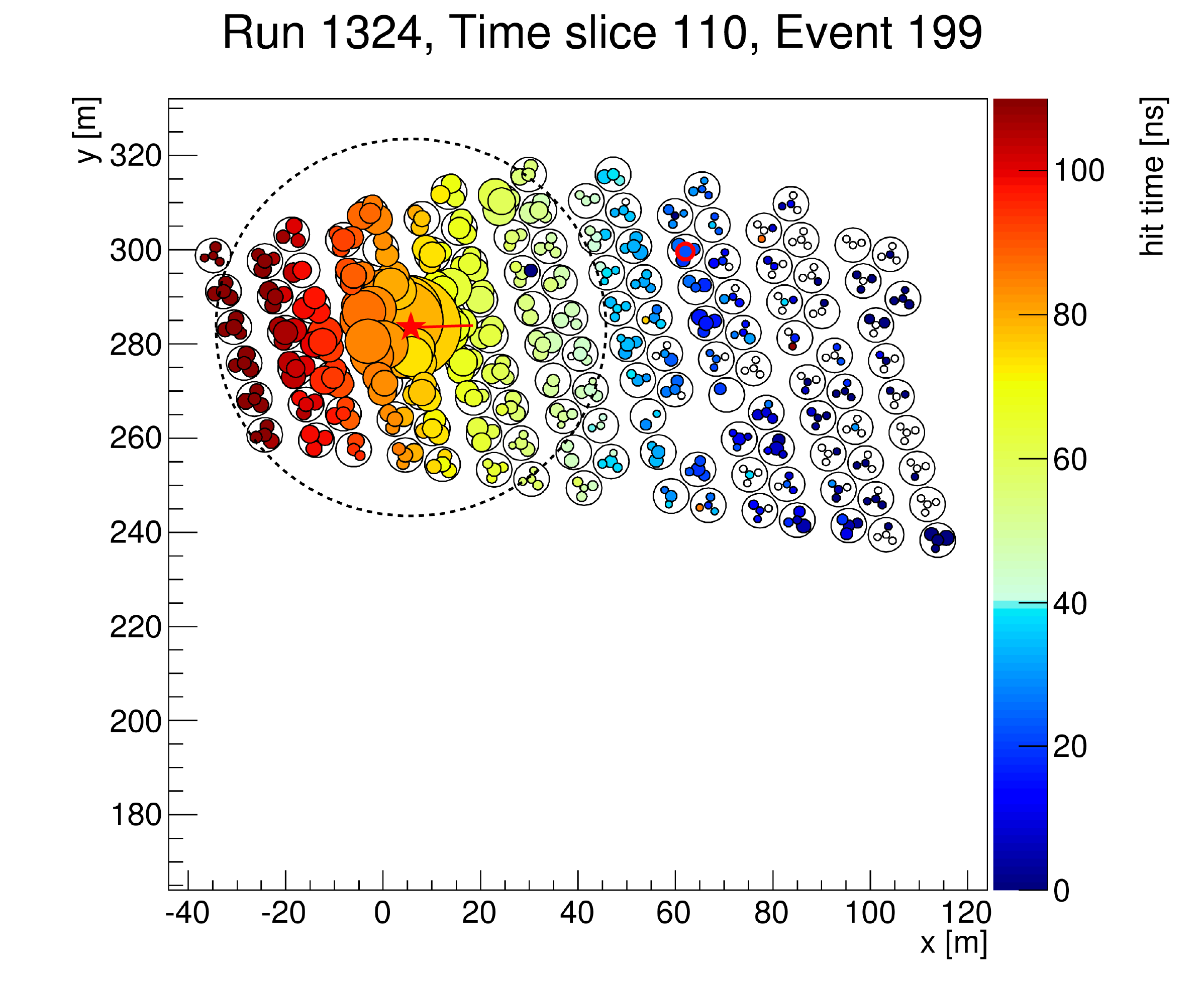}
\includegraphics[width=0.5\textwidth]{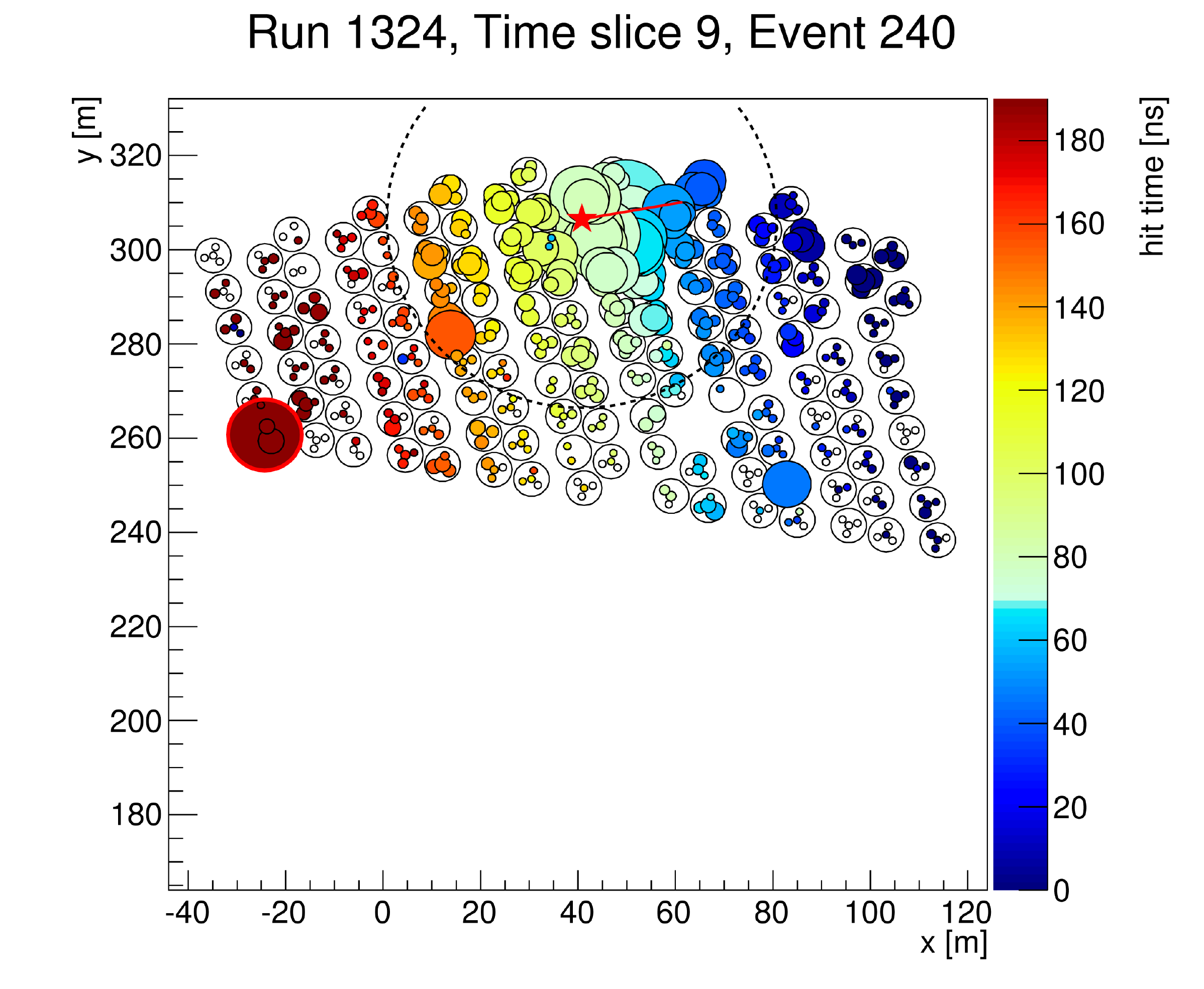}
\caption{A gamma-like event (left) and a hadron-like event (right) observed by the Pass 1 configuration.  Each filled circle is a PMT signal, with the color indicating the hit time and the size indicating the charge.  The reconstructed shower core is marked with a red star and a dashed circle indicates the 40\,m radius around the reconstructed core.  The highlighted red circle represents the location of the PMT with the maximum observed PE outside of the 40\,m radius from the core.}
\label{fig:gh}
\end{figure}

The primary background of this analysis is air showers induced by hadrons. The shape of the lateral distribution of gamma-induced air showers differs from that of hadron-induced showers. The NKG function used for fitting the shower core describes the lateral distributions of gamma-ray showers.  Consequently, the chi-square values resulting of the core fit are on average smaller for gamma-ray showers than for cosmic-ray showers with the same number of triggered PMTs, and this parameter is used for background discrimination. Furthermore, hadronic showers produce pions with large transverse momenta that decay to gamma rays and muons.  These subshowers are likely to produce large signals in PMTs far from the shower core. Fig.\,\ref{fig:gh} shows a typical gamma-like event and a hadron-like event observed with the Pass 1 configuration. The ratio of the number of triggered PMTs to the number of photoelectrons (PE) in the PMT that detects the strongest signal outside of a radius of 40\,m from the shower core is found to result in good gamma/hadron separation performance. A third selection applied to the reduced chi-square distribution of the shower angle fit removes poorly reconstructed air showers.  All cuts are optimized by maximizing the sensitivity of the Pass 1 data to emission from the Crab Nebula, the brightest steady source at TeV. In the lowest \textit{f} bin, about 79\% of gamma ray events and 38\% of cosmic ray events pass the cuts, whereas for the highest \textit{f} bin these numbers drop to about 13\% and 0.03\% respectively.

\subsection{Sky Maps and Maximum Likelihood Method}
The arrival directions of air showers that pass the cuts are binned in a signal map in equatorial coordinates using HEALPIX \citep{gorski2005}, a pixelization scheme that splits the unit sphere into twelve equal-area tessellations.  Each of the twelve tessellations are subdivided into an $N_\mathrm{side} \times N_\mathrm{side}$ grid, giving 12\,$N^2_\mathrm{side}$ pixels.  In this analysis, $N_\mathrm{side}$ is chosen to be 512, which provides an average pixel size of $0.11\degree$.  The bin size is much smaller than the typical point response of the Pass 1 detector, discussed in Section 3.2.

The background at each location in the sky is estimated from the data using the direct integration method described in \citet{atkins2003}.  It is computed by the convolution of the local event hour angle and declination distribution with the all-sky event rate recorded during a predefined integration step.  An integration step of two hours is used in this analysis, corresponding to $30\degree$ in right ascension, to emphasize structures smaller than this angular scale.

To convert the gamma ray counts to flux, a simulation-based model of the detector response to sources of gamma rays is used. The detector response file describes the energy distribution and the point spread function (PSF) as a function of source declination and fraction hit \textit{f} bins. It is generated with the air shower simulation program CORSIKA \citep{heck1998} and the detector simulation package GEANT4 \citep{agostinelli2003}.  Since the data are subdivided into three epochs reflecting the size of the Pass 1 array, three sets of detector simulations with 365, 426, and 480 PMTs were generated.

The first step of the maximum likelihood fit is to build a source model characterized by the position and spectrum of the source.  Right ascension $\alpha$ and declination $\delta$ are used to describe the source position.  The source spectrum in the present analysis is assumed to obey a simple power law 

\begin{equation}
\frac{dN}{dE} = I_0 \left(\frac{E}{E_0}\right)^{-\Gamma}, 
\end{equation}

\noindent where $I_0$ is the differential flux normalization, $E_0$ is the pivot energy, which is chosen at where the differential flux normalization is least dependent on the spectral index, and $\Gamma$ is the spectral index. The study presented here assumes a fixed index of 2.3 due to the limited sensitivity in this data set. The index of 2.3 is representative of measured values for known Galactic objects. 

The region of interest (ROI) used for a likelihood fit needs to be larger than the angular resolution of the detector to include most photons from a given source. However, it is not always possible to find an ROI with photons from one single source due to a high potential of source confusion in the Pass 1 Galactic plane data.  Therefore, the source model may need to contain more than one source and in this case the expected event count becomes

\begin{equation}
\lambda_{ij} = b_{ij} + \sum_{k}\gamma_{ijk},
\end{equation}

\noindent where $b_{ij}$ is the background events in the $j$th pixel of the $i$th \textit{f} bin, and $\gamma_{ijk}$ is the expected number of gamma rays from the $k$th source in the $j$th pixel of $i$th \textit{f} bin.  The event count is convolved with the detector response.  As the observed event count in each pixel is distributed according to a Poisson distribution, the probability of observing $N$ number of events given an expected count $\lambda$ from the source model is 

\begin{equation}
P(N;\lambda) = \frac{\lambda^N e^{-\lambda}}{N!}.
\end{equation}

The likelihood given a parameter set $\vv\theta = (\alpha,\delta,I_0)$ in the source model is the product of the likelihood of each pixel in an ROI and in each \textit{f} bin:

\begin{equation}
\mathcal{L}(\vv{\theta} | \vv{N}) = \prod_{i}^{\mathrm{f\,bins}}\prod_{j}^{\mathrm{ROI}} P(N_{ij};\lambda_{ij}).
\end{equation}

\noindent where $N_{ij}$ and $\lambda_{ij}$ are the observed and expected event count in the $j$th pixel of the $i$th \textit{f} bin, respectively.  The logarithm of the likelihood is used for ease of computation:

\begin{equation}
\mathrm{ln}\,\mathcal{L}(\vv{\theta} | \vv{N}) = \sum_{i}^{\mathrm{f\,bins}}\sum_{j}^{\mathrm{ROI}}(N_{ij} \mathrm{ln}\,\lambda_{ij}-\lambda_{ij}),
\label{equ:logL}
\end{equation}

\noindent The term $N_{ij}!$ is discarded from Eq.\,\ref{equ:logL} since it is independent of the parameters in the source model.  The log likelihood is maximized with respect to the parameter set $\vv{\theta}$ in the source model using the MINUIT package \citep{minuit}.

A likelihood ratio test is performed to decide how many sources are needed to properly model an ROI. To decide if the one-source model is preferred over the background-only model, the log likelihood of the background-only model ln\,$\mathcal{L}_0$ is computed first.  Then the log likelihood ln\,$\mathcal{L}_1$ of the one-source model is computed.  The test statistic ($TS$) defined by

\begin{equation}
TS = -2(\mathrm{ln}\,\mathcal{L}_0-\mathrm{ln}\,\mathcal{L}_1)
\end{equation}

\noindent is used to compare the goodness of fit between the two models. The $TS$-value is converted to a p-value, which is the probability of the data being consistent with the background-only hypothesis. The same likelihood ratio test can be used to compare between two models with N and N+1 sources.

In this iterative process, an additional source characterized by three free parameters (right ascension, declination, and differential flux normalization) is added to the model at each step, with the positions and amplitudes of the existing sources free to change in response to the new source.  We repeat this procedure as long as $\Delta TS>15$ after adding a new source, corresponding to a p-value of 1\%. The $\Delta TS$ threshold was chosen \textit{a priori}. After the iteration in each ROI, a simultaneous fit of the differential source fluxes in all regions is performed while the source positions are fixed. This is done to take into account photons from a source with centroid position just outside of an ROI but still contributing to the ROI due to the PSF of the detector. Finally, the differential flux and $TS$ value of each source are obtained by fitting a single source while treating other sources as part of the background with fixed positions and differential fluxes.

\begin{figure}
\centering
\includegraphics[width=\textwidth]{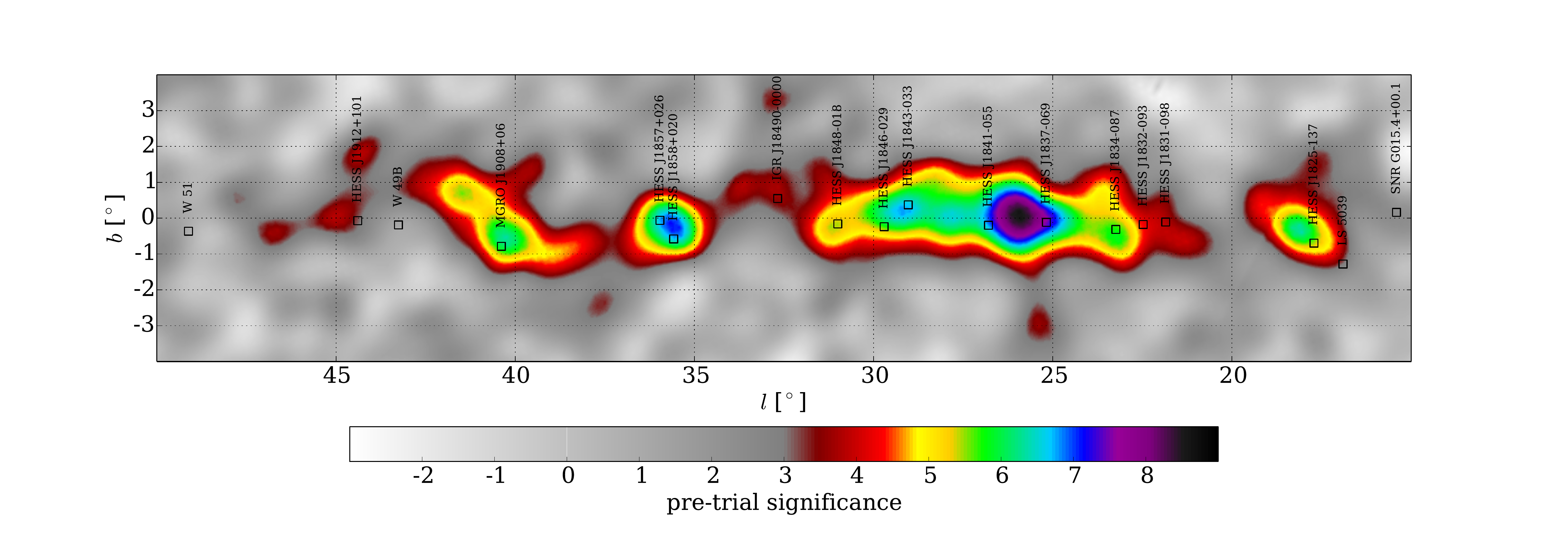}
\includegraphics[width=\textwidth]{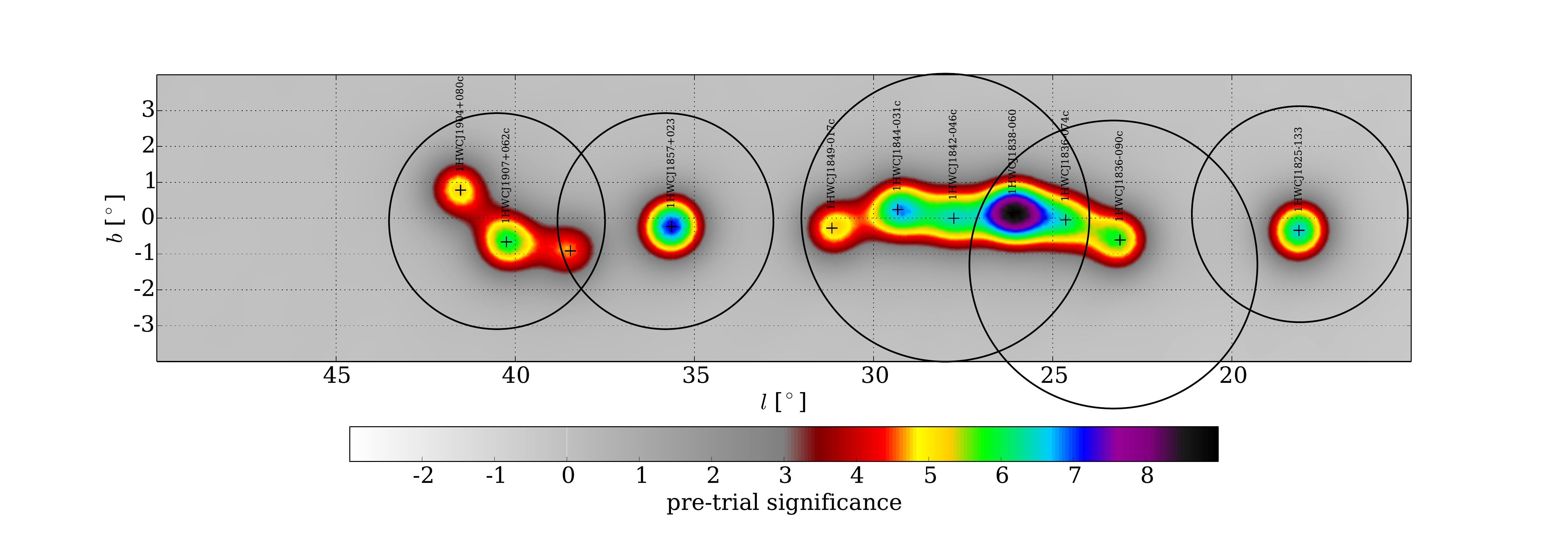}
\includegraphics[width=\textwidth]{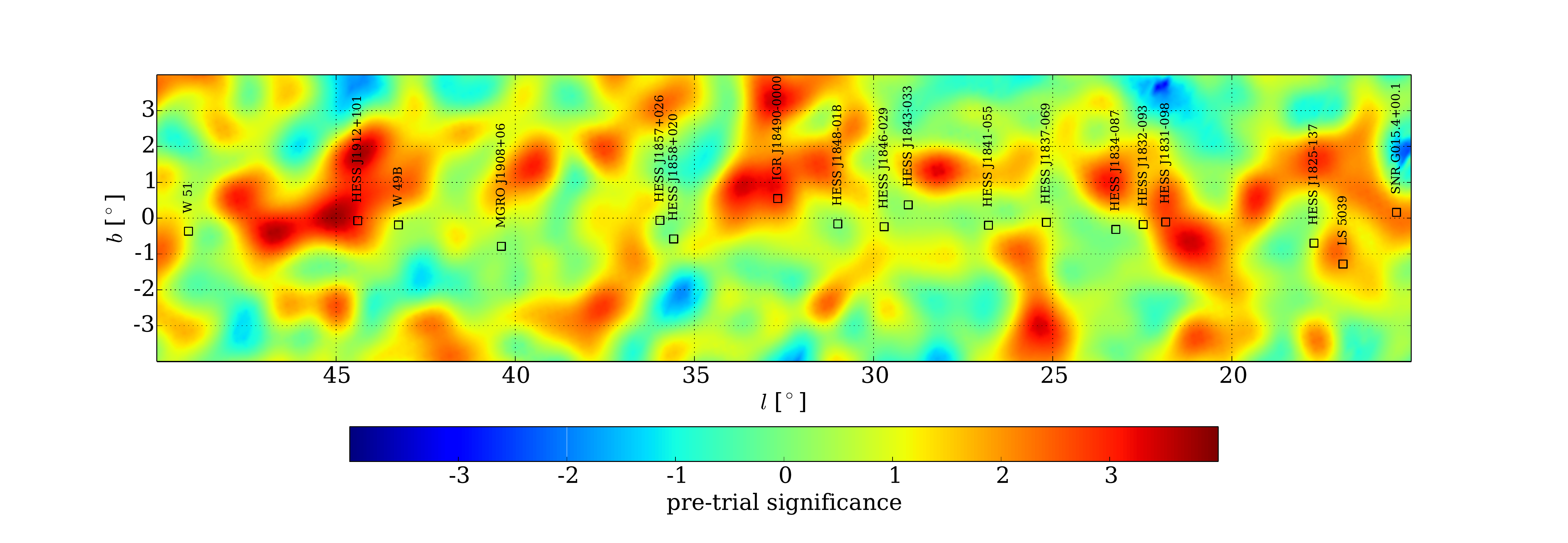}
\caption{\textit{Top:} Significance map;
\textit{Middle:} Model with 11 seed sources (crosses) and no uniform surface brightness fit.  Black circles indicate the 5 regions of interest;
\textit{Bottom:} Residual significance map.  The open squares mark TeV sources in \citet{tevcat}.}
\label{fig:skymap}
\end{figure}

Fig.\,\ref{fig:skymap} shows the pre-trials significance map of the inner Galaxy region surveyed in this paper. The map is made by moving a putative point source through each pixel, performing a maximum likelihood fit of differential flux normalization with the spectral index fixed at 2.3, and converting the $TS$ value to a significance according to Wilks' theorem ($\sigma=\sqrt{TS}$ with 1 degree of freedom (DoF) for this map).  There are multiple $>5\,\sigma$ hotspots in the surveyed region as shown in Fig.\,\ref{fig:skymap}.  The area is divided into five ROIs as defined in Table\,\ref{table:srcROI}. In each of these ROIs, the procedure described above is applied to simultaneously account for the flux contributions of neighboring sources.

\begin{deluxetable}{cccc}
\tablewidth{0pt}
\tablecaption{Definition of the 5 regions of interest}
\tablehead{
\colhead{Region} & \colhead{RA} & \colhead{Dec} & \colhead{Radius}}
\startdata
1 & $286.4\degree$ & $6.7\degree$ & $3.0\degree$ \\
2 & $284.2\degree$ & $2.5\degree$ & $3.0\degree$ \\
3 & $280.7\degree$ & $-4.1\degree$ & $4.0\degree$ \\
4 & $279.6\degree$ & $-9.2\degree$ & $4.0\degree$ \\
5 & $275.9\degree$ & $-13.1\degree$ & $3.0\degree$ \\
\enddata
\label{table:srcROI}
\end{deluxetable}

\subsection{Systematic Uncertainty and Simulation Studies}

\begin{figure}
\centering
\epsscale{.80}
\plotone{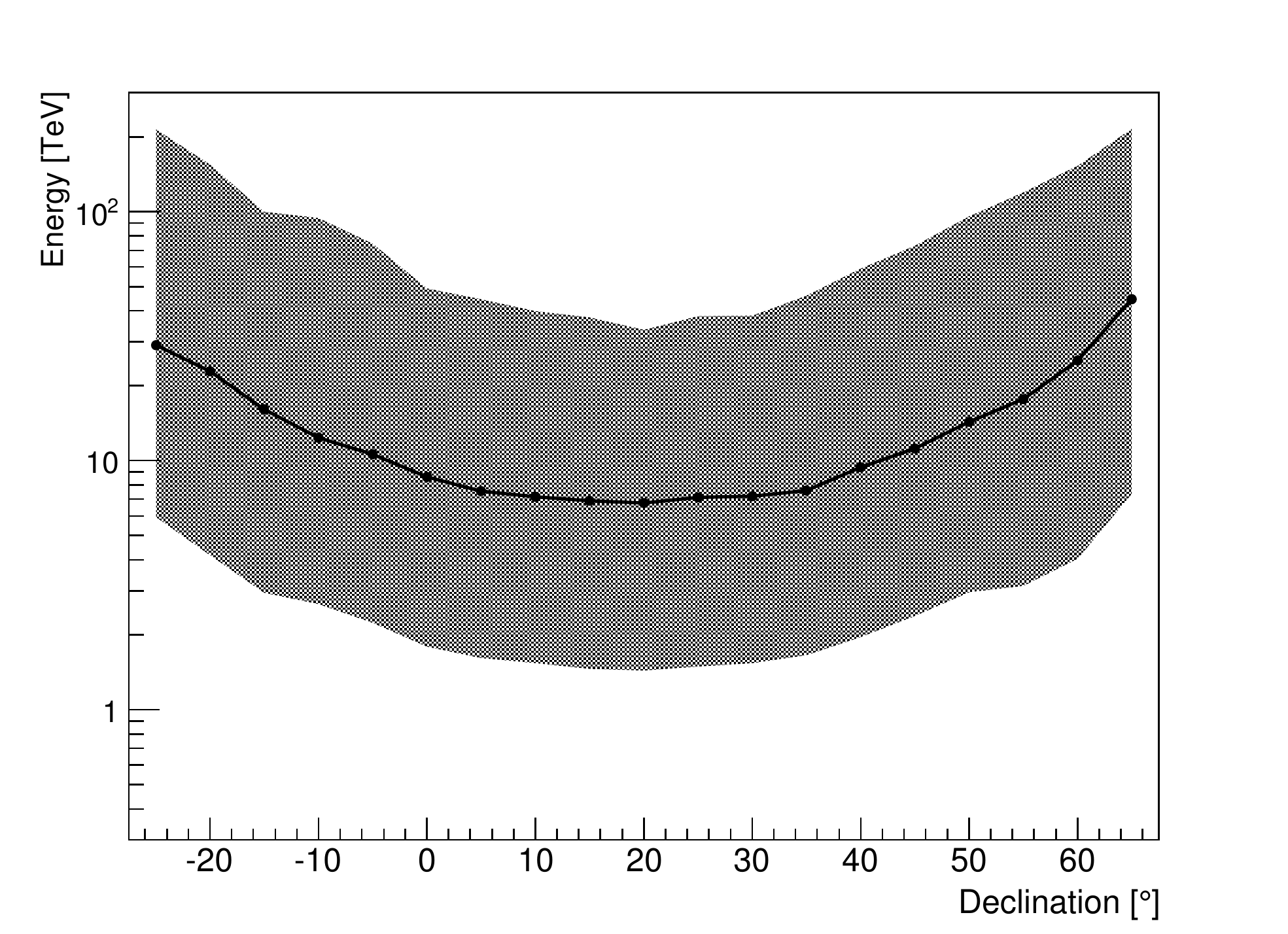}
\caption{Median gamma-ray energy of the Pass 1 data as a function of declination along with the 10\% and 90\% quantiles, derived from simulations and assuming a spectral index of -2.3.}
\label{fig:energy}
\end{figure}

The energy range in which the present analysis is most sensitive is estimated using detector and air shower simulations.  Since more inclined air showers travel through more atmosphere, the median gamma-ray energy of the Pass 1 data is a function of declination.  The declination dependence is shown in Fig.\,\ref{fig:energy} for the spectral index assumption $\Gamma=2.3$.  The median energy increases from 7\,TeV for a declination of $+19\degree$ to $\sim30$\,TeV for declinations of $-26\degree$ and $+64\degree$ assuming a spectral index of 2.3.

\begin{figure}
\centering
\epsscale{.80}
\plotone{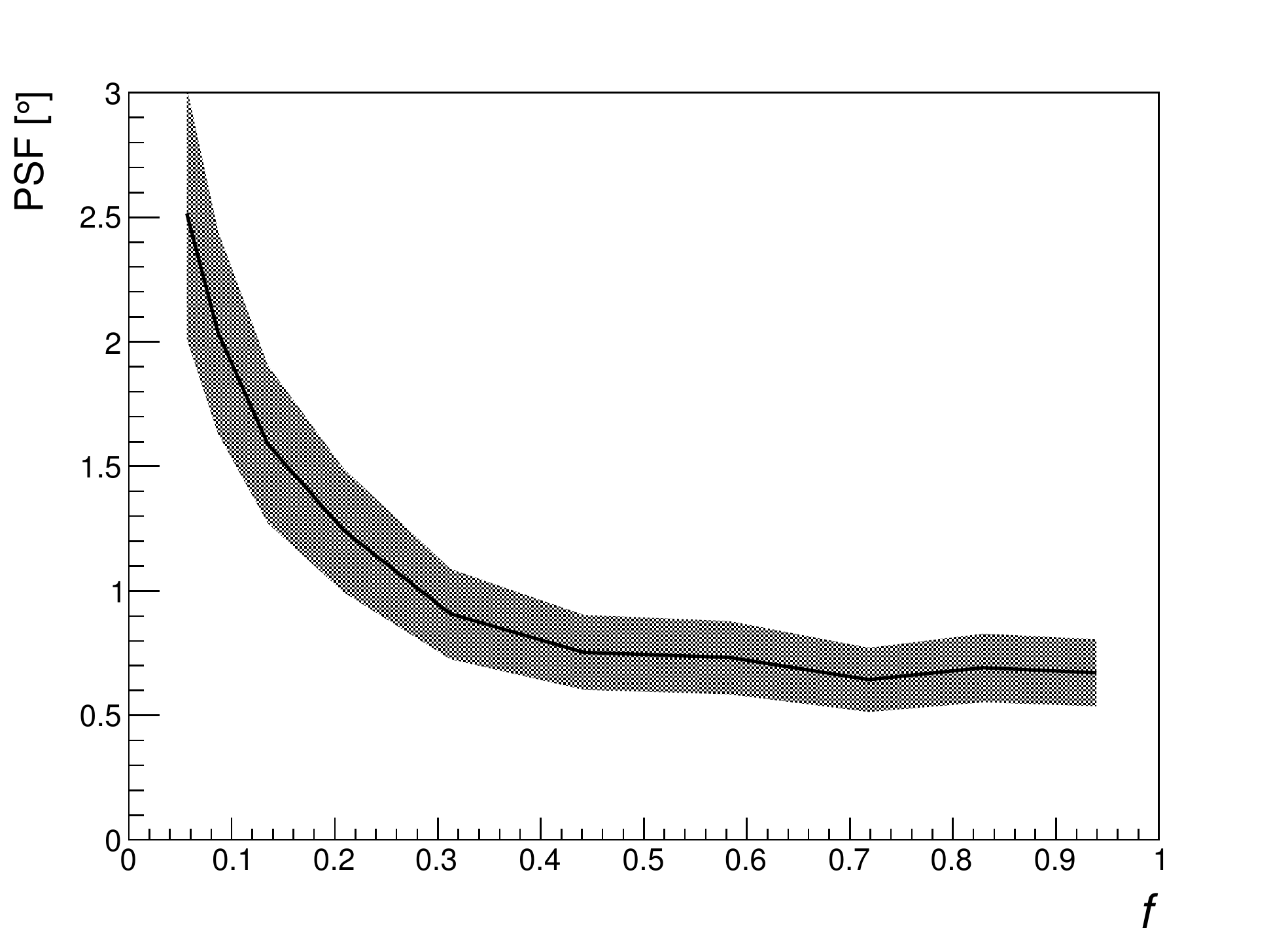}
\caption{PSF measured in data from the Crab Nebula (68\% containment) as a function of the fraction \textit{f} of PMTs used in event reconstruction. The systematic uncertainty on the measured PSF (gray) is $<20\%$.  A linear combination of two Gaussians is needed to correctly characterize the PSF.}
\label{fig:psf}
\end{figure}

The Crab Nebula is used to measure the point spread function in the data since its extent is much smaller than the PSF of the HAWC detector. We describe the PSF of the Pass 1 detector as the linear combination of two Gaussian functions. Fitting the emission from the Crab Nebula region in each \textit{f} bin, the 68\% containment region is found to vary from $2.5\degree$ to $0.6\degree$ depending on the event size that correlates with the fractional number of PMTs in the event (see Fig.\,\ref{fig:psf} on 68\% PSF as a function of \textit{f} bin). The PSF measured in the data from the Crab Nebula is used for all other sources in this likelihood analysis.

\begin{figure}[h]
\centering
\epsscale{1.0}
\plottwo{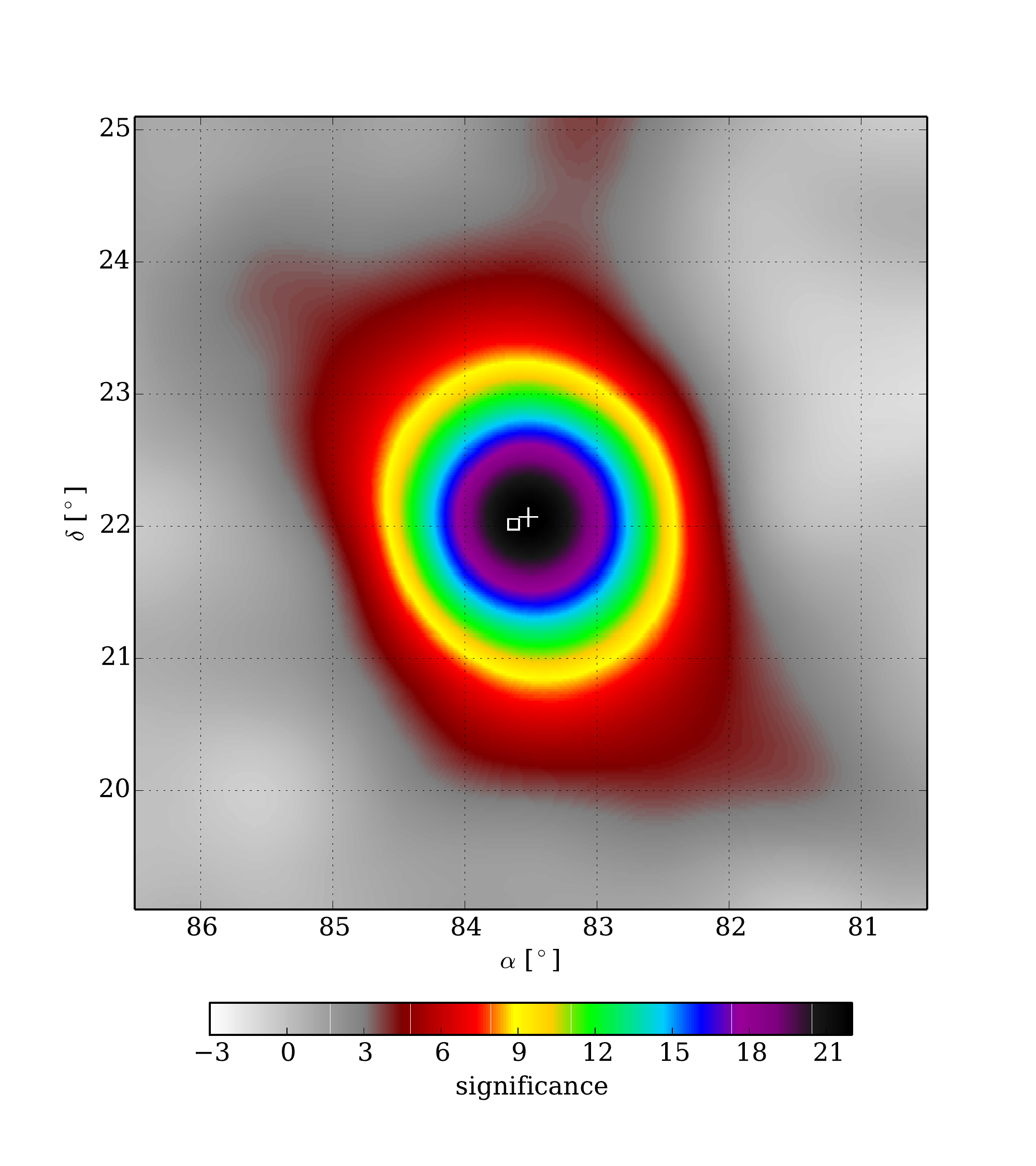}{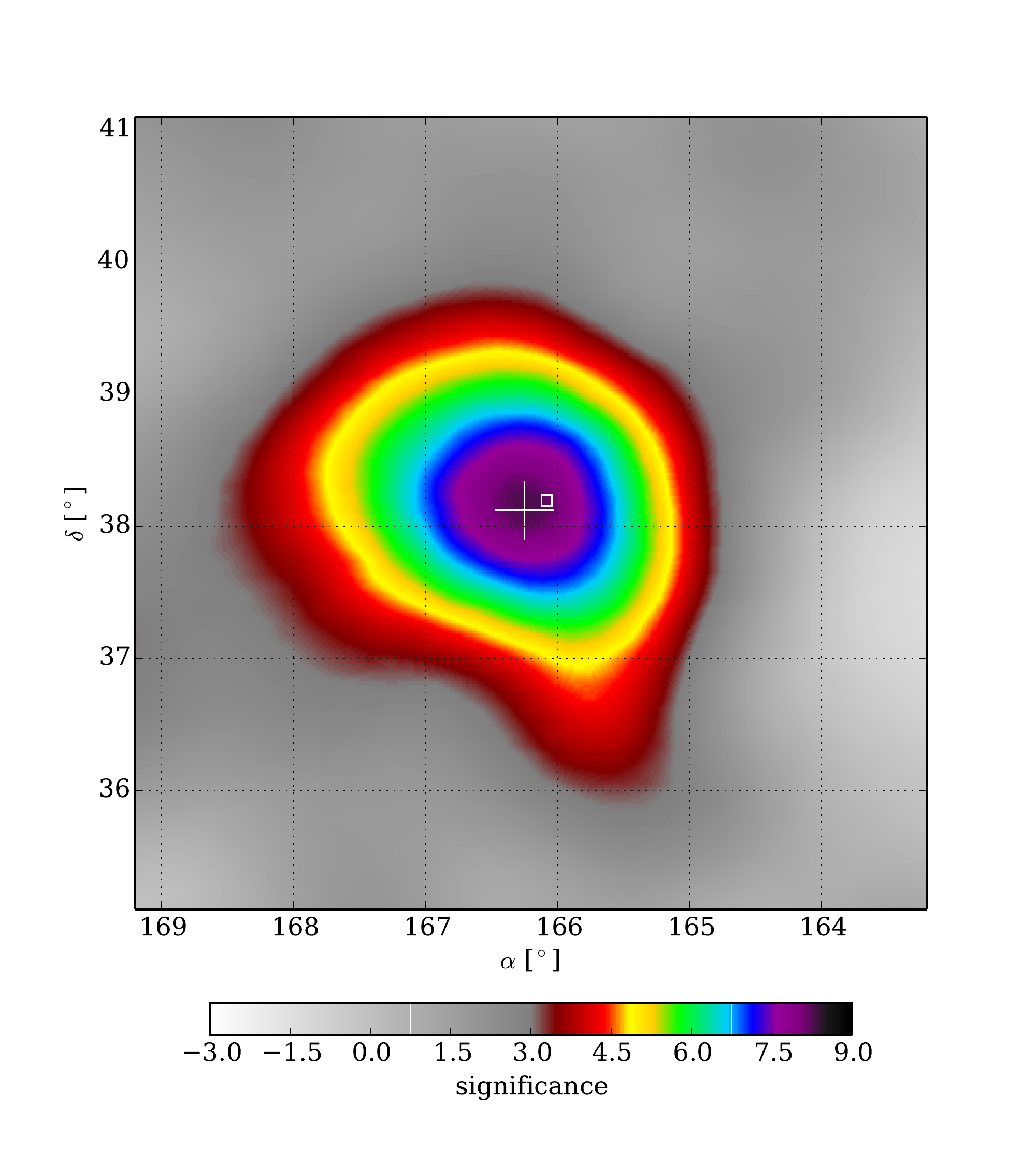}
\caption{Significance maps of the Crab Nebula (left) and Mrk 421 (right).  The white squares indicate the source positions from \citet{tevcat} and the white crosses are measured positions in the Pass 1 data along with $1\,\sigma$ errors.
}
\label{fig:crabMrk421}
\end{figure}

The pointing of the detector is verified using data from the Crab Nebula and the blazar Markarian (Mrk) 421.  Both source regions are relatively isolated in the sky and do not suffer overlap with additional sources.  A likelihood fit with right ascension, declination, and differential flux normalization is performed using a power law spectral assumption with the fixed spectral index of 2.6 for the Crab Nebula and 3.0 for Mrk 421.  The results are summarized in Table\,\ref{table:pointing}.  The Crab Nebula spectral index assumption is chosen based on previous IACT measurements, and for Mrk 421 a soft spectral index of 3.0 is chosen due to known spectral cutoff that is not modelled here.  The positions of both sources in the Pass 1 data are consistent with measurements by IACTs (for example \citet{aharonian2004,albert2007}).  Changing the spectral index used in the fit of the Crab Nebula between 2.0 and 3.0 shifts the best fit position by $<0.07\degree$.  The significance maps in the vicinity of the Crab Nebula and Mrk 421 are shown in Fig.\,\ref{fig:crabMrk421}.  These maps are made by moving a putative point source through each pixel and performing a maximum likelihood fit of the differential flux normalization with spectral index fixed at 2.6 and 3.0, respectively.  Then the $TS$ value in each pixel is converted to significance according to Wilks' theorem.

\begin{deluxetable}{ccccccc}
  \tablewidth{0pt}
  \tablecaption{Pointing in J2000}
  \tablehead{
    \multirow{2}{*}{Source} & \multicolumn{2}{c}{SIMBAD Database} & \multicolumn{4}{c}{Pass 1} \\
      &RA ($\degree$) & Dec ($\degree$) & RA\tablenotemark{a} ($\degree$) & Dec\tablenotemark{a} ($\degree$) & \it{TS} & significance
  }
  \startdata
    Crab & 83.63 & 22.01 & $83.53\pm0.06$ & $22.06\pm0.06$ & 491.4 & $22.2\,\sigma$\\
    Mrk 421 & 166.11 & 38.21 & $166.22\pm0.18$ & $38.14\pm0.18$ & 69.0 & $7.8\,\sigma$
  \enddata
  \label{table:pointing}
\end{deluxetable}

The Crab Nebula is the most significant source in the Pass 1 data.  The spectrum of this source has been well measured by IACTs.  The Pass 1 data cover a similar gamma-ray energy range as the IACT data.
The pivot energy for the Crab Nebula analysis is selected to be 4\,TeV in order to minimize the dependence of the differential flux normalization on the spectral index in the fit.  It differs from the pivot energy of 1\,TeV used by Whipple, HEGRA, H.E.S.S., and VERITAS and 0.3\,TeV used by MAGIC, so the IACT differential flux normalizations at 4\,TeV are computed from the respective flux normalizations and spectral indices \citep{whippleCrab,aharonian2004,hessCrab,veriCrab,magicCrab}.
Table\,\ref{table:crabflux} shows the Pass 1 differential flux normalization from the Crab Nebula with a spectral index assumption of 2.3, which is used for the analysis of the Inner Galaxy region, as well as the differential flux normalization with a spectral index assumption of 2.6, which is close to the index measured by IACTs. The differential flux of the Crab Nebula as measured in the Pass 1 data is within 15\% of the IACT measurements.  The flux of Mrk 421 is beyond the scope of this paper due to known variability of the blazar during the period covered by this dataset and will be presented in detail in an upcoming publication.

\begin{deluxetable}{cccc}
\tablewidth{0pt}
\tablecaption{Differential Flux Normalization Comparison of the Crab Nebula Assuming a Simple Power Law\label{table:crabflux}}
\tablehead{
\colhead{Instrument} & \colhead{Flux Normalization at 4\,TeV} & \colhead{Reported Flux Normalization} & \colhead{Spectral Index\tablenotemark{a}}\\
\colhead{} & \colhead{($10^{-13}\,\text{TeV}^{-1} \text{cm}^{-2} \text{s}^{-1}$)} & \colhead{ at 1\,TeV\tablenotemark{a}($10^{-11}\,\text{TeV}^{-1} \text{cm}^{-2} \text{s}^{-1}$)} & \colhead{}
}
\startdata
HAWC Pass 1 & $8.57 \pm 0.45$ & - & 2.30\tablenotemark{b} \\
HAWC Pass 1 & $8.25 \pm 0.40$ & - & 2.60\tablenotemark{b} \\
Whipple & $10.1$ & $3.2\pm0.17\pm0.6$ & $2.49 \pm 0.06\pm0.04$ \\
HEGRA & $7.49$ & $2.83\pm0.04\pm0.6$ &$2.62 \pm 0.02\pm0.05$ \\
H.E.S.S. & $9.00$ & $3.45\pm0.05\pm0.7$ & $2.63 \pm 0.01\pm0.09$ \\ 
MAGIC & $9.25$ & $57\pm2\pm6$\tablenotemark{c} &$2.48 \pm 0.03\pm0.2$ \\
VERITAS & 8.83 &$3.48\pm0.14 \pm 1.08$ & $2.65 \pm 0.04 \pm 0.3$
\enddata
\tablenotetext{a} {The first quoted uncertainty is statistical uncertainty and the second is systematic uncertainty.}
\tablenotetext{b} {Assumed spectral index.}
\tablenotetext{c} {At 300 GeV.}
\end{deluxetable}

The determination of the differential flux normalization in Pass 1 depends on the assumption of the spectral index, but the variance can be minimized by selecting the best pivot energy. For example, the derived Crab differential flux normalization increases by 5\% if an index of 2.3 is chosen instead of 2.6 for a pivot energy of 4\,TeV, while a pivot energy of 9\,TeV results in a 30\% increase in flux normalization.  For the sources in the region of the Galactic plane analyzed here a spectral index of 2.3 is assumed and the best pivot energy varies from 4\,TeV to 9\,TeV depending on declination. In order to minimize the variance in the flux normalization determination due to different spectral indices, different pivot energies are used for sources as a function of declination.  According to simulations, this leads to a systematic change of $-10\%$ to $+20\%$ in the fitted differential flux for a source spectral index ranging between 2.1 and 2.8.

There are three other major contributions to the systematic error of the flux normalization that have been studied using data and simulations:

\begin{enumerate}
  \item Detector configuration variability.  The number of active PMTs changed continuously in this dataset but only three configurations are modeled with the detector simulations.  The event passing rate is found to vary $<20\%$ among the three simulated configurations, which is equivalent to the resulting uncertainty on the flux estimate.
  \item Angular resolution.  The measured PSF on the Crab Nebula is used to compute the expected number of gamma rays in each pixel in a given model.  The error in the measured PSF width from the Crab Nebula is $<20\%$, which corresponds to 15-20\% uncertainty in the flux estimate. 
  \item Charge scale.  The core fitter and gamma/hadron cuts rely on the charge measurements by PMTs, which are based on the ToT-charge calibration.  Up to 20\% change in the flux estimate is observed due to the uncertainty in the charge scale estimated in studies of single muons with HAWC.
\end{enumerate}

Each source of systematic uncertainties contributes $\sim20\%$. In addition, a minor contribution of $\sim8\%$ from the uncertainty of atmospheric modeling is taken into account \citep{Dirk2015}. We add systematic uncertainties in quadrature for a total systematic uncertainty in the flux normalization of $\sim40\%$.

\section{Results}
\begin{deluxetable}{ccccccccccc}
\tabletypesize{\scriptsize}
\rotate
\tablewidth{0pt}
\tablecaption{Galactic Source Detections and Candidates}
\tablehead{
\colhead{Region} & \colhead{Source\tablenotemark{a}} & \colhead{$\Delta TS$\tablenotemark{b}} & \colhead{RA ($\degree$)\tablenotemark{c}} & \colhead{Dec ($\degree$)\tablenotemark{c}} & \colhead{l ($\degree$)\tablenotemark{c}} & \colhead{b ($\degree$)} & \colhead{Differential Flux (Pivot Energy)} & \colhead{$TS$\tablenotemark{c}} & \colhead{Post-trials} \\
\colhead{} & \colhead{} & \colhead{} & \colhead{} & \colhead{} & \colhead{} & \colhead{} & \colhead{($10^{-14} \text{TeV}^{-1} \text{cm}^{-2} \text{s}^{-1}$)} & \colhead{} & \colhead{Significance}
}
\startdata
1 & 1HWC\,J1907+062c & 40.9 & 286.8$\pm$0.2 & 6.2$\pm$0.2 & 40.2$\pm$0.2 & -0.7$\pm$0.2 & 22.0$\pm$4.6 (4\,TeV) & 32.8 & $4.6\,\sigma$ \\
& 1HWC\,J1904+080c & 26.8 & 286.1$\pm$0.2 & 8.0$\pm$0.2 & 41.5$\pm$0.2 & 0.8$\pm$0.2 & 19.0$\pm$4.4 (4\,TeV) & 26.5 & $3.9\,\sigma$ \\
& -- & 16.2 & 286.2$\pm$0.4 & 4.5$\pm$0.3 & 38.5$\pm$0.4 & -0.9$\pm$0.4 & N/A & 17.2 & $2.5\,\sigma$ \\ \hline
2 & 1HWC\,J1857+023 & 52.1 & 284.3$\pm$0.2 & 2.3$\pm$0.2 & 35.6$\pm$0.2 & -0.2$\pm$0.2 & 18.0$\pm$3.0 (5\,TeV) & 50.2 & $6.2\,\sigma$ \\ \hline
3 & 1HWC\,J1838-060 & 74.7 & 279.6$\pm$0.3 & -6.0$\pm$0.2 & 26.1$\pm$0.3 & 0.2$\pm$0.3 & 11.3$\pm$1.2 (7\,TeV) & 48.9 & $6.1\,\sigma$ \\
& 1HWC\,J1844-031c & 47.4 & 281.0$\pm$0.2 & -3.1$\pm$0.2 & 29.3$\pm$0.2 & 0.2$\pm$0.2 & 11.8$\pm$2.4 (6\,TeV) & 33.7 & $4.7\,\sigma$ \\
& 1HWC\,J1849-017c & 25.2 & 282.3$\pm$0.3 & -1.7$\pm$0.2 & 31.2$\pm$0.3 & -0.3$\pm$0.3 & 9.1$\pm$2.2 (6\,TeV) & 24.9 & $3.7\,\sigma$ \\
& 1HWC\,J1842-046c & 23.7 & 280.5$\pm$0.3 & -4.6$\pm$0.3 & 27.8$\pm$0.3 & 0.0$\pm$0.3 & 7.0$\pm$1.6 (7\,TeV) & 23.2 & $3.4\,\sigma$ \\ \hline
4 & -- & 70.7 & 279.7$\pm$0.2 & -6.1$\pm$0.3 & 26.1$\pm$0.3 & 0.0$\pm$0.3 & 11.3$\pm$1.2 (7\,TeV) & 48.9 & same source as J1838-060 \\
& 1HWC\,J1836-090c & 33.6 & 278.9$\pm$0.3 & -9.0$\pm$0.2 & 23.1$\pm$0.3 & -0.6$\pm$0.3 & 5.8$\pm$1.3 (8\,TeV) & 26.6 & $3.9\,\sigma$ \\
& 1HWC\,J1836-074c & 18.4 & 279.1$\pm$0.3 & -7.4$\pm$0.3 & 24.6$\pm$0.3 & 0.0$\pm$0.3 & 6.9$\pm$1.4 (7\,TeV) & 22.0 & $3.2\,\sigma$ \\ \hline
5 & 1HWC\,J1825-133 & 40.8 & 276.3$\pm$0.1 & -13.3$\pm$0.2 & 18.1$\pm$0.2 & -0.3$\pm$0.2 & 7.3$\pm$1.4 (9\,TeV) & 40.6 & $5.4\,\sigma$ \\
\enddata
\tablenotetext{a}{
   $\mathrm{L}$ist of Galactic source detections and candidates.  The positions reported here are in epoch J2000, and the differential flux normalization assumes a spectral index of 2.3.  Only statistical uncertainties are quoted in this table.}
\tablenotetext{b}{
  $\Delta TS$ of a model with one more source ($\Delta$DoF=3) over the previous model.}
\tablenotetext{c}{
  $TS$ of a source over the background model while treating other sources as part of the background.}
\label{table:source}
\end{deluxetable}

Table\,\ref{table:source} lists the epoch J2000 positions, differential flux normalizations, $TS$, and the post-trials significances of the detections and candidates from this analysis.  Eleven seed sources are initially identified with the $\Delta TS>15$ criterion and are used in the source model. To calculate the number of trials within the inner Galaxy search region of 280 square degrees, a Monte Carlo study of fluctuations based on the measured background was performed to obtain the p-value distribution of the maximum significance on background-only maps within the inner Galaxy region. A fit to the corresponding tail distribution has been performed, and the exponent derived from this fit is equal to the number of trials. The resulting number of trials is $424 \pm 3$. After accounting for trials \citep{trials}, ten source detections and candidates remain with $>3\,\sigma$. Table\,\ref{table:counterpart} lists the possible counterparts of each source and differential flux normalization comparison with known TeV sources.  The discussion is separated into TeV source detections and source candidates below using a criterion of $5\,\sigma$ significance after trials.

\begin{deluxetable}{ccccccccc}
\tabletypesize{\scriptsize}
\rotate
\tablewidth{0pt}
\tablecaption{Possible TeV Gamma-Ray Source Counterparts}
\tablehead{
\colhead{Source} & \colhead{Possible Counterpart} & Counterpart & \colhead{Distance to} & \colhead{Published Angular} & \colhead{Extrapolated} & \colhead{Flux} & \colhead{Pivot Energy}\\
\colhead{} & \colhead{} & \colhead{Classification} & \colhead{Counterpart ($\degree$)} & \colhead{Extent ($\degree$)} & \colhead{Published Flux}  & \colhead{Normalization\tablenotemark{a}} & \colhead{(TeV)} 
}
\startdata
1HWC\,J1907+062c & MGRO\,J1908+06 & UID & 0.38 & $<2.6$ & 36 & 22.0$\pm$4.6 & 4\\
 & HESS\,J1908+063 & UID & 0.19 & $0.34^{+0.04}_{-0.03}$ & 22.5 & \\
 & MGRO\,J1908+06 (ARGO) & UID & 0.29 & $0.49 \pm 0.22$ & 61 & \\
 & MGRO\,J1908+06 (VERITAS) & UID & 0.04 & $0.44 \pm 0.02$ & 20.0 & \\ \hline
1HWC\,J1857+023 & HESS\,J1857+026 & UID & 0.37 & $(0.11\pm0.08) \times (0.08\pm0.03)$ & 13.0 & 18.0$\pm$3.0 & 5\\
 & MAGIC\,J1857.2+0263 & PWN & 0.33 & $(0.17\pm0.03) \times (0.06\pm0.03)$ & 16.6 & \\ 
 & HESS\,J1858+020 & UID & 0.35 & $(0.08\pm0.02) \times (0.02\pm0.04)$ & 1.8 & \\ \hline 
1HWC\,J1838-060 & HESS\,J1841-055 (ARGO) & UID & 0.16 & $0.40^{+0.32}_{-0.22}$ & 41 & 11.3$\pm$1.2 & 7 \\
 & HESS\,J1841-055 & UID & 0.77 & $(0.41\pm0.04) \times (0.25\pm0.02)$ & 11.7 & \\ 
 & HESS\,J1837-069 & PWN & 0.97 & $(0.12\pm0.02) \times (0.05\pm0.02)$ & 6.1 & \\ \hline 
1HWC\,J1844-031c & HESS\,J1843-033 & UID & 0.32 & extended & N/A & 11.8$\pm$2.4 & 6 \\
 & HESS\,J1846-029 & PWN & 0.61 & point-like & 1.1 & \\ 
 & ARGO\,J1841-0332 & UID & 0.87 & point-like & N/A & \\ \hline
1HWC\,J1849-017c & HESS\,J1848-018 & MSC\tablenotemark{b} & 0.20 & $0.32\pm0.02$ & 2.5 & 9.1$\pm$2.2 & 6 \\ \hline
1HWC\,J1836-090c & HESS\,J1834-087 & UID & 0.31 & point-like$+ (0.17\pm0.01)$ & 1.0 & 5.8$\pm$1.3 & 8 \\ 
 & HESS\,J1834-087 (MAGIC) & UID & 0.41 & $0.14\pm0.04$ & 2.0 \\ \hline 
1HWC\,J1836-074c & HESS\,J1837-069 & PWN & 0.55 & $(0.12\pm0.02)\times(0.05\pm0.02)$ & 6.1 & 6.9$\pm$1.4 & 7 \\ \hline
1HWC\,J1825-133 & HESS\,J1825-137 & PWN & 0.55 & $(0.23\pm0.02)\times(0.26\pm0.02)$ & 10.6 & 7.3$\pm$1.4 & 9\\ \hline
\enddata
\tablenotetext{a}{
  The differential flux normalization in units of $10^{-14} \text{TeV}^{-1} \text{cm}^{-2} \text{s}^{-1}$, assumes a spectral index of 2.3.  Only statistical uncertainties are quoted here.}
\tablenotetext{b}{
  Massive Star Cluster}
\label{table:counterpart}
\end{deluxetable}

After the identification of 11 seed sources, a model and a uniform surface brightness for the entire region was fitted simultaneously to the data.  While for the sources a spectral index of 2.3 is assumed, the spectral index assumption for the uniform surface brightness is 2.5.  The $\Delta TS$ of adding the uniform surface brightness as another free parameter to the source model is 33, i.e. $5.7\,\sigma$ that the uniform surface brightness component is preferred.  The fitted surface brightness at 5\,TeV is $(1.6 \pm 0.4) \times 10^{-11}\,\text{TeV}^{-1} \text{cm}^{-2} \text{s}^{-1} \text{sr}^{-1}$, which is compatible with the average diffuse flux of $(1.0 \pm 0.2) \times 10^{-11}\,\text{TeV}^{-1} \text{cm}^{-2} \text{s}^{-1} \text{sr}^{-1}$ reported by H.E.S.S. extrapolated to 5\,TeV within the same region \citep{hessdiffuse}.  However, the uniform surface brightness measured in this dataset is not simply diffuse emission but also a combination of unidentified sources, source extensions, and photon contaminations from sources due to uncertainties in the PSF.  As evident in the residual map of Fig\,\ref{fig:skymap}, there are several $3\,\sigma$ regions around known TeV sources that are not detected in this dataset and are contributing to this uniform surface brightness fit. The uniform surface brightness fit is not included in Table\,\ref{table:source}.  The contribution to the source differential flux normalization is $<30\%$ of the smallest reported flux. 

\subsection{Source Detections}
The source 1HWC\,J1857+023 is detected at $6.2\,\sigma$ post trials and is $\sim 0.4\degree$ away from both HESS\,J1857+026 and HESS\,J1858+020.  These two TeV sources were discovered by the H.E.S.S. collaboration during their Galactic plane survey and are $\sim 0.7\degree$ apart.  The flux of HESS\,J1857+026 is approximately an order of magnitude higher than HESS\,J1858+020 \citep{hessUID}, and the differential flux normalization from 1HWC\,J1857+023 is compatible with the combined flux of HESS\,J1857+026 and HESS\,J1858+020.  Both of the HESS sources were detected as extended, with HESS\,J1857+026 as the larger of the two.  \citet{magic1857} reported energy dependent morphology for HESS\,J1857+026 with two distinct components, MAGIC\,J1857.2+0263 and MAGIC\,J1857.6+0297.  These sources cannot be resolved with these data from the partial HAWC array.  The spectrum reported by MAGIC is for the entire region and is compatible with the differential flux normalization derived from this dataset.  

1HWC\,J1838-060 is detected at $6.1\,\sigma$ post trials and is located in the middle of the known TeV sources HESS\,J1837-069 and HESS\,J1841-055.  This detection overlaps with the extension of HESS\,J1841-055, and the differential flux normalization is compatible with that reported by H.E.S.S.\citep{hessUID}.  \citet{argo1841} reported a $5.3\,\sigma$ detection by ARGO near the position of HESS\,J1841-055, with a $0.4\degree$ source extent and is closest to this detection.  The ARGO source extends towards HESS\,J1837-069 and includes several \textit{Fermi}-LAT sources.  The flux reported by ARGO, when converted to differential flux at 7\,TeV for comparison, is $\sim 4\times$ the differential flux normalization derived from this dataset.

1HWC\,J1825-133 has a post-trials significance of $5.4\,\sigma$.  It is $\sim 0.5\degree$ to the south of the HESS\,J1825-137 centroid position, which is an extended PWN with spectral softening as a function of distance from the pulsar towards a southeast direction \citep{hess1825}.  The simple power-law flux derived from this dataset is lower than the flux extrapolated from the simple power-law assumption measured by H.E.S.S.  However, \citet{hess1825} reported the spectrum is unlikely to be a simple power law and presented several alternative fits.  The derived flux normalization from this dataset is most compatible with the energy dependent photon index power law fit by H.E.S.S.  There is also a nearby PSR $\sim 0.4\degree$ away, PSR\,J1826-1256, seen by \textit{Fermi}-LAT \citep{3fgl} and associated with the Eel nebula \citep{snrcat}.

\subsection{Source Candidates} 
1HWC\,J1907+062c is $4.6\,\sigma$ post trials with a best-fit position that is compatible with previously reported positions of MGRO\,J1908+06 (see \citet{milagroSurvey,hess1908,argo1908,veri1908} for example).  The differential flux normalization is consistent with the flux measured by H.E.S.S. and VERITAS and in agreement with Milagro given the statistical uncertainties of both instruments.  \citet{veri1908} reported strong excess near the pulsar PSR\,J1907+0602 but also extends toward SNR G\,40.5-0.5.  The Pass 1 dataset is not able to resolve the spatial morphology of this source.

1HWC\,J1904+080c has a post-trials significance of $3.9\,\sigma$. There is currently no previously reported TeV detection near this location.  The nearest gamma-ray source is 3FGL\,J1904.9+0818 at $0.3\degree$ away \citep{3fgl}. However, this is a weak detection from the \textit{Fermi}-LAT 3FGL catalog, at $<5\,\sigma$, with no known association. 

1HWC\,J1844-031c has a post-trials significance of $4.7\,\sigma$ and is spatially coincident with HESS\,J1843-033, which is classified as an unidentified source \citep{hess1843}.  However, the morphology of this detection appears to extend towards HESS\,J1846-029, a pulsar wind nebula \citep{hess1846}.  \citet{argoSurvey} reported a $4.2\,\sigma$ excess, ARGO\,J1841-0332, associated with HESS\,J1843-033 despite being $0.7\degree$ away, due to the large systematic pointing error at high zenith angle.

1HWC\,J1849-017c is detected at $3.7\,\sigma$ post trials and is positionally coincident with the extended source HESS\,J1848-018, which is possibly associated with the star forming region W43 \citep{hess1848}. The differential flux normalization at 6\,TeV from this dataset is $\sim 3.5\times$ the flux reported by H.E.S.S.  \citet{hess1848} reported an index of 2.8 and the spectral index assumption of 2.3 in this analysis would result in a different flux normalization by 20\%.  
More importantly, diffuse emission from this star forming region that contains a molecular cloud could contribute more to the differential flux normalization derived from the Pass 1 dataset than that measured by IACTs, which have a smaller angular integration region.

1HWC\,J1842-046c has a post-trials significance of $3.4\,\sigma$ and has no clear gamma-ray association.  There is an X-ray source, SNR G\,27.4+0.0 (Kes 73), located  $\sim0.4\degree$ \citep{snrcat} away from this candidate.

1HWC\,J1836-090c is detected at $3.9\,\sigma$ post trials.  It is spatially coincident with HESS\,J1834-087 and the SNR W41 \citep{HESSsurvey,magic1834,hess1834}.  The differential flux normalization from the Pass 1 dataset at 8\,TeV is $\sim 6\times$ higher than the flux reported in \citet{hess1834}.  The source is reported by H.E.S.S. as having a central point-like component and an extended component.  A similarly extended component is also seen by \textit{Fermi}-LAT.  The region contains a candidate pulsar at the center of the SNR W41, and two scenarios are supported by \citet{hess1834}: PWN or SNR interaction with a nearby molecular cloud.  The cloud density traced by $^{13}$CO appears wider than the H.E.S.S. detection and may contribute to the increased flux seen in the Pass 1 data with HAWC.

1HWC\,J1836-074c has a post-trials significance of $3.2\,\sigma$, with the nearest TeV PWN, HESS\,J1837-069 \citep{HESSsurvey}, $\sim 0.5\degree$ away with a compatible differential flux.  There is also a GeV source 3FGL\,J1837.6-0717 \citep{3fgl} that is $\sim 0.3\degree$ away with no association.

\section{Conclusion}
A survey of the inner Galaxy has been presented in the region of Galactic longitude $l\in[15\degree, 50\degree]$ and latitude $b\in[-4\degree, + 4\degree]$ using 283 days of data from the partial HAWC Gamma-Ray Observatory from August 2013 to July 2014.  Three sources have been detected at $>5\,\sigma$ with an additional seven candidate sources detected at $>3\,\sigma$ after accounting for trials. While associations with previously published IACT detections are not always within the experimental uncertainties, about half of them have differential flux normalizations that are compatible with the previous detections.  

A likelihood method similar to the \textit{Fermi}-LAT source-finding algorithm has been applied to data from an extended air shower array for the first time to properly address challenges arising from source identification.  The point sources presented here have differential fluxes $>20\%$ of the Crab Nebula flux at several TeV.  Future analyses will build on this method to most effectively exploit the increased sensitivity and improved pointing of data collected with the full HAWC array, and to explore other regions of the sky visible to the observatory.

\acknowledgements
We acknowledge the support from:
the US National Science Foundation (NSF);
the US Department of Energy Office of High-Energy Physics;
the Laboratory Directed Research and Development (LDRD)
program of Los Alamos National Laboratory;
Consejo Nacional de Ciencia y Tecnolog\'{\i}a (CONACyT), Mexico (grants 55155, 105666, 122331, 132197);
Red de F\'{\i}sica de Altas Energ\'{\i}as, Mexico;
DGAPA-UNAM (grants IG100414-3, IN108713, IN121309, IN115409, IN113612);
VIEP-BUAP (grant 161-EXC-2011);
the University of Wisconsin Alumni Research Foundation;
the Institute of Geophysics, Planetary Physics, and
Signatures at Los Alamos National Laboratory;
the Luc Binette Foundation UNAM Postdoctoral Fellowship program.

Facilities: \facility{HAWC}

\end{document}